\documentclass[%
 reprint,
 amsmath,amssymb,
 aps,
]{revtex4-2}

\usepackage{graphicx}
\usepackage{dcolumn}
\usepackage{bm}

\usepackage[ruled,linesnumbered]{algorithm2e}
\usepackage{ragged2e}
\usepackage{color}
\usepackage{xcolor}
\usepackage{array}
\usepackage{amsmath}
\usepackage{amsthm}
\usepackage{amssymb}

\begin{document}

\preprint{APS/123-QED}

\title{Machine learning of network inference enhancement from noisy measurements}%
\author{Kai Wu}
\affiliation{School of Artificial Intelligence, Xidian University, Xi'an 710071, China 
}%

\author{Yuanyuan Li}
\affiliation{%
Guangzhou Institute of Technology, Xidian University, Guangzhou 510555, China
}%


\author{Jing Liu}
\affiliation{%
 Guangzhou Institute of Technology, Xidian University, Guangzhou 510555, China 
}


\date{\today}

\begin{abstract}
Inferring networks from observed time series data presents a clear glimpse into the interconnections among nodes. Network inference models, when dealing with real-world open cases, especially in the presence of observational noise, experience a sharp decline in performance, significantly undermining their practical applicability. We find that in real-world scenarios, noisy samples cause parameter updates in network inference models to deviate from the correct direction, leading to a degradation in performance. Here, we present an elegant and efficient model-agnostic framework tailored to amplify the capabilities of model-based and model-free network inference models for real-world cases. 
Extensive experiments across nonlinear dynamics, evolutionary games, and epidemic spreading, showcases substantial performance augmentation under varied noise types, particularly thriving in scenarios enriched with clean samples.
\end{abstract}

\maketitle

Complex networks \cite{1}, \cite{2} play a pivotal role in comprehending, identifying, and regulating complex systems \cite{3} across diverse domains such as biology \cite{4,5,6}, physics \cite{7}, and technology \cite{8}. In reality, direct access to network structures remains limited, often necessitating inference from measurement data \cite{9,cliff2023unifying}. These inference methodologies can be categorized into two primary groups: model-free and model-based. Most model-free techniques establish links between nodes by detecting statistical dependencies within time series data 
\cite{10,11,12,13,14,15,16}. 
On the other hand, model-based strategies, such as stability selection \cite{17}, compressed sensing \cite{7,18,19,20}, Bayesian inference \cite{21,22,23}, online learning \cite{24}, evolutionary computation \cite{25}, presuppose a priori knowledge of system behaviors. Yet, the efficacy of existing network inference models experiences significant degradation in the presence of observational noise 
\cite{26,27}. While certain studies have explored the resilience of their network inference methodologies against minor noise levels \cite{7,18,19,28,29,30,31,32,33}, they have primarily focused on enhancing individual method, lacking broader applicability to other network inference techniques.

This article presents a model-agnostic framework, named MANIE (Model-Agnostic Network Inference Enhancement), designed to enhance exsiting network inference methods across both model-based and model-free paradigms when confronted with measurement data polluted by observational noise (not dynamical noise \cite{banerjee2019using,prill2015noise,lipinski2015using}). The presence of noise can divert the optimization process of network inference models from the optimal trajectory exhibited by clean samples. Our approach aims to address this challenge by attenuating the influence of noisy samples on network inference models. Leveraging curriculum learning techniques \cite{35,36}, we progressively identify and downweight noisy samples within the network inference process. The simplicity of the MANIE framework lies in its reliance on weighted samples, enabling seamless integration into existing network inference models. Remarkably, MANIE offers broader support for diverse network inference methodologies compared to existing strategies. Notably, its efficiency shines through, yielding significant performance improvements across a spectrum of noisy scenarios.

\begin{figure*}[htbp]
\includegraphics[width=0.8\linewidth]{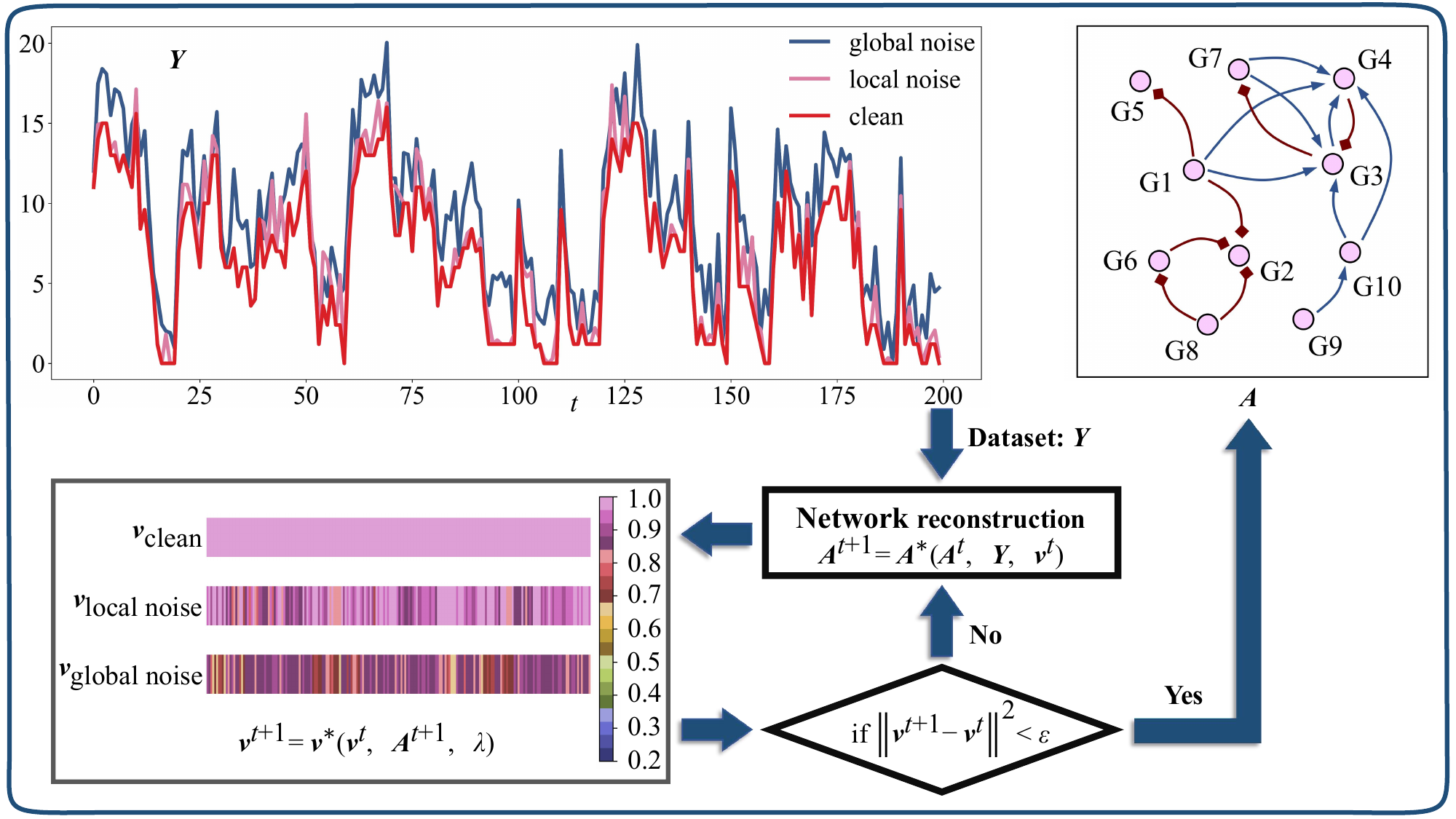}
\caption{Graphical illustration of MANIE framework based on noisy time series. $\bm{A}^\ast$ or $\bm{v}^\ast$ represents the value of $\bm{A}$ or $\bm{v}$ when the loss function takes its minimum. “global noise” refers to all samples being contaminated by noise, while “local noise” indicates that only a portion of the samples is contaminated. “clean” refers to untainted sample data.}
\label{fig1}
\end{figure*}

%
We consider a system governed by the differential equation:
\begin{equation}
\label{eq1}
\dot{\bm{x}}=\mathbf{F}\left(\bm{A},\bm{x}\left(t\right)\right)+\bm{\xi}(t)
\end{equation}
where $\bm{x}\left(t\right)=\left[x_1\left(t\right),\cdots,x_N\left(t\right)\right] \in \mathbb{R}^N$ is the state of units, $\dot{\bm{x}}=d\bm{x}\left(t\right)/dt$ denotes its temporal derivative, $\bm{A} \in \mathbb{R}^{N\times N}$ denotes the network structure among units, $\mathbf{F}:\ \mathbb{R}^N\rightarrow\mathbb{R}^N$ stands for an unknown nonlinear function that indicates direct interactions among those units, and $\bm{\xi}\left(t\right)=\left[\xi_1\left(t\right),\cdots,\xi_N\left(t\right)\right] \in \mathbb{R}^N$ defines the external noise. We aims to deduce $\bm{A}$ from periodic or non-periodic dynamics $\bm{Y}=\left[\bm{x}_1\left(1\right),\cdots,\bm{x}_1\left(M\right);\cdots;\bm{x}_N\left(1\right),\cdots,\bm{x}_N\left(M\right)\right]+\bm{\epsilon}$, where $\bm{\epsilon}\in \mathbb{R}^{N\times M}$ is observational noise and $M$ represents the sample size.

Figure \ref{fig1} demonstrates the graphical illustration of MANIE framework.
We cast the problem of model-based network inference enhancement into the following form
\begin{eqnarray}
\label{eq4}
\min_{\bm{A}}\sum_{t=1}^Mv_t\mathcal{L}(\bm{Y}_t, \bm{F}, \bm{A})+\frac{1}{2}\lambda\sum_{t=1}^{M}{(v_t^2-2v_t)}
\end{eqnarray}
where the loss function $\mathcal{L}$ measures the disparity between the simulated data by network inference method and the observed data. The loss function for model-free methods is: $\sum_{t=1}^Mv_t\mathcal{L}(\bm{Y}_t, \bm{A})$. Any network inference method with the above optimization objectives can be embedded in MANIE. $v_t \in \left[0,\ 1\right]$ allows for adjusting the contribution of $\bm{Y}_t$ to $\mathcal{L}$. The left term $\frac{1}{2}\lambda\sum_{t=1}^{M}{(v_t^2-2v_t)}$ \cite{43,44} indicates that the model inclines to select clean samples (with smaller losses) instead of noisy samples (with larger losses) (see \cite{citation-key} for more information). When the parameter $\lambda$ becomes larger, it tends to incorporate more, probably noisy, samples to obtain $\bm{A}$. Its convexity ensures the soundness of this term for optimization.

We jointly learn $\bm{A}$ and $\bm{v}=\{v_1, \cdots, v_M\}$ by the alternative optimization strategy (AOS) \cite{44} with the gradually increasing $\lambda$. This iterative process continues until convergence ($\bm{v}$ hardly changes), yielding the final target $\bm{A}$. For simplicity, we demonstrate the AOS process using the model-free case as an example.

\begin{enumerate}
\item Given $\bm{v}^i$ and $\bm{A}^i$, we employ one network inference method to obtain $\bm{A}^{i+\mathbf{1}}$. 
\begin{equation}
\label{eq6}
\bm{A}^{i+\mathbf{1}} = \arg \min_{}\sum_{t=1}^M v_t^i\mathcal{L}(\bm{Y}_t, \bm{A}^i)
\end{equation}
\item Calculate $\bm{v}$ by solving the following problem under fixed $\bm{A}^{i+\mathbf{1}}$, $\bm{v}^i$, and $\lambda$, readjusting the contribution of each sample to $\mathcal{L}$.
\begin{equation}
\label{eq5}
{v_t}^{i+1}=\arg\min_{v_t^i\in\left[0,1\right]}v_t^i\mathcal{L}(\bm{Y}_t, \bm{A}^{i+\mathbf{1}})+\frac{1}{2}\lambda\sum_{t=1}^{M}{(v_t^2-2v_t)},
\end{equation}
$\frac{1}{2}\lambda\sum_{t=1}^{M}{(v_t^2-2v_t)}$ is a convex function of $\bm{v}$, and thus the global minimum can be obtained at $\frac{\partial\left(v_t^i\mathcal{L}\left(\bm{Y}_t, \bm{A}^{i+\mathbf{1}})+\frac{1}{2}\lambda\sum_{t=1}^{M}{(v_t^2-2v_t)}\right)\right)}{\partial v_t^i}=\mathcal{L}(\bm{Y}_t, \bm{A}^{i+\mathbf{1}})+\lambda\left(v_t^i-1\right)=0$. Thus, the corresponding closed-formed solution is
\begin{eqnarray}
v_t^i = 
\left\{
\begin{array}{l}
1-\mathcal{L}(\bm{Y}_t, \bm{A}^{i+\mathbf{1}})/\lambda, \ if \mathcal{L}(\bm{Y}_t, \bm{A}^{i+\mathbf{1}})<\lambda,  \\
0,   \ \  \ \ \ \ \ \ \ \ \ \ \ \ \ \ \ \ \ \ \ \ \ \ \ \ \ \ \   otherwise
\end{array}\right.
\label{eq7}
\end{eqnarray} 
\end{enumerate}

We also update $\lambda=c\lambda$, where $c$ can be selected from $\{1.05, 1.25, 1.5\}$.
We adopted the area under the receiver-operating-characteristic curve (AUC) score as the metric, with a higher AUC score denoting superior inference performance. The benchmark used in this paper, along with their corresponding network structures, are derived from existing literature on network inference—specifically, the datasets used in the embedded network inference methods referenced in the literature. The details of experimental setting is shown in \cite{citation-key}.

\begin{figure*}[htbp]
\includegraphics[width=1\linewidth]{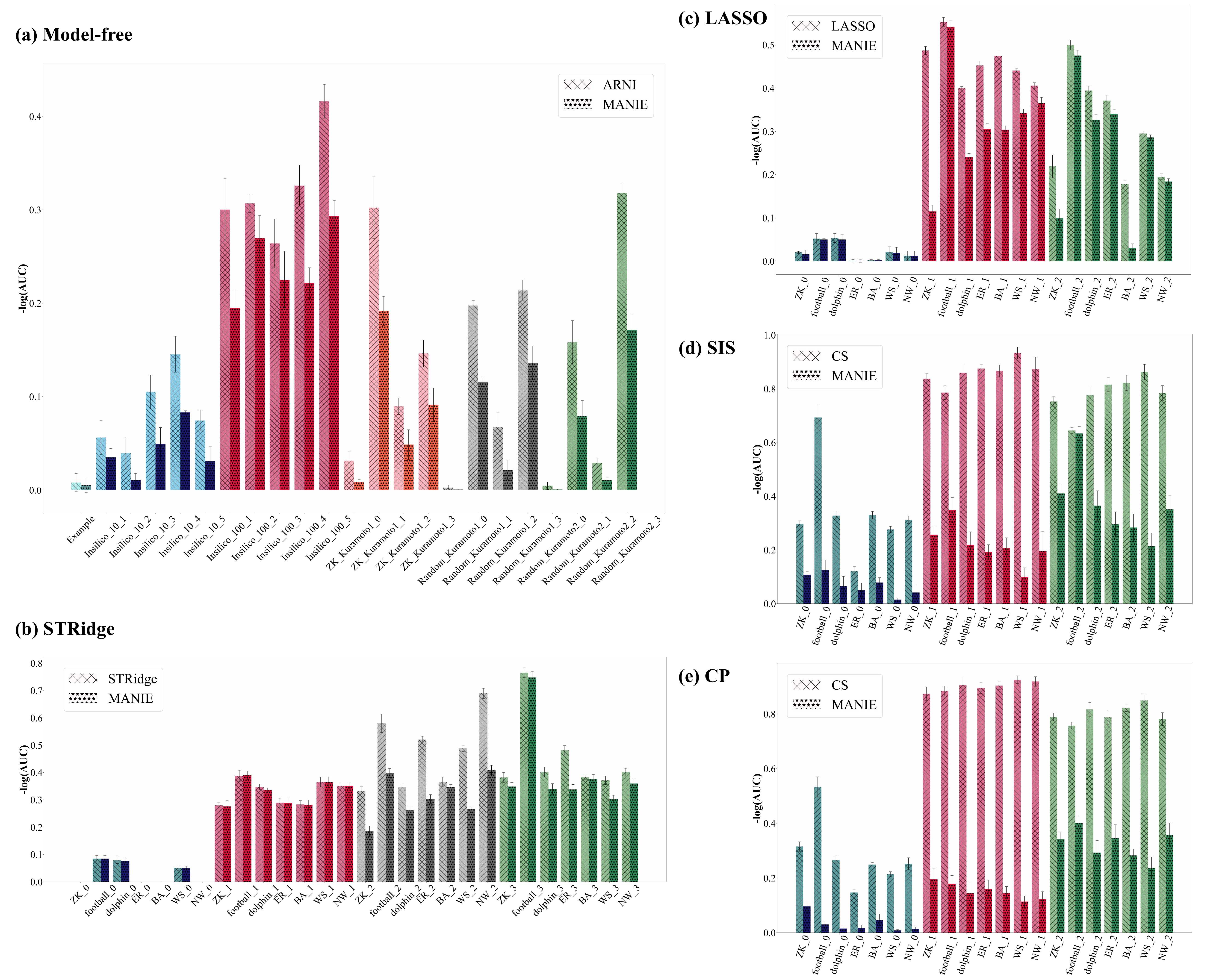}
\caption{AUC scores for different methods for various scenarios. The negative logarithm is taken on the AUC in order to convert it into an unrestricted indicator with a more intuitive interpretation. Smaller \(-\log_{2}(\text{AUC})\) indicates superior inference performance. (a) Enhancing Model-free network inference methods. "Example" is a 5$\times$5 example network, the specific form of which is described in the supplementary material. Insilico\_10 and Insilico\_100 pertain to networks of sizes 10 and 100 in DREAM4 challenge \cite{5}, where the suffix represents different networks. ZK\_Kuramoto1 illustrates the dynamics of Zachary’ s karate club network (ZK) \cite{45} under a Kuramoto-1 oscillator. Similarly, Random\_Kuramoto1 and Random\_Kuramoto2 depict the dynamics of directed random networks under Kuramoto-1 and Kuramoto-2 oscillators \cite{46}, respectively. The term "\_0" denotes noise-free time series. The suffix "\_1" indicates the addition of white Gaussian noise with a signal-to-noise ratio (SNR) of 10 to the global time series. For "\_2", white Gaussian noise is added at each timestep with a probability of 0.5 and an SNR of 10 dB. In "\_3", white Gaussian noise is added at each timestep with a probability of 0.5 and an SNR ranging from 0 to 10, randomly generated. Enhancing model-based inference from EG data using MANIE with embedded (b) STRidge or (c) LASSO. Here, we employ ZK, football \cite{57}, and dolphin \cite{58} networks. Additionally, we utilize four synthetic networks with 40 nodes each: Erdős-Rényi random network (ER) \cite{53}, Barabási-Albert scale-free network (BA) \cite{54}, Newman-Watts small-world network (NW) \cite{55}, and Watts-Strogatz small-world network (WS) \cite{56}. For these cases, (i) "\_1": random noise with an amplitude of 10 added to each time step's recordings, with a probability of 0.5; (ii) "\_2": random noise with random amplitudes less than 10, added to each time step's recordings with a probability of 0.5; (iii) "\_3": random noise added to the global data, magnitudes set at 1, 5, and 10 respectively, with the final outcomes averaged. Enhancing model-based inference of propagation networks for (d) SIS dynamics or (e) CP dynamics using MANIE with embedded CS. We introduce two common types of noise: (i) "\_1", where 10\% of nodes in the binary time series are missing; (ii) "\_2", wherein each record is misremembered with a probability of 0.1, causing a reversal between 0 and 1.
}
\label{fig2}
\end{figure*}
\textit{Enhancing model-based network inference}. 
Following model-based network inference studies \cite{18,19,20}, we infer networks from evolutionary game (EG) models \cite{50,51}, the classic susceptible-infected-susceptible dynamics (SIS) \cite{48}, and contact processes (CP) \cite{49}.

Following \cite{18,20}, LASSO \cite{39} and sequential threshold ridge regression (STRidge) \cite{40} are embedded in MANIE. The experimental findings on the dynamics of EG are presented in Figure \ref{fig2}(b)(c). 
In noise-free scenarios, MANIE is at least as capable as STRidge and LASSO, and in some cases, even exhibits a slight advantage over them. When noise is present, MANIE consistently achieves greater performance improvement compared to LASSO and STRidge. 

\begin{figure*}[htbp]
\includegraphics[width=1\linewidth]{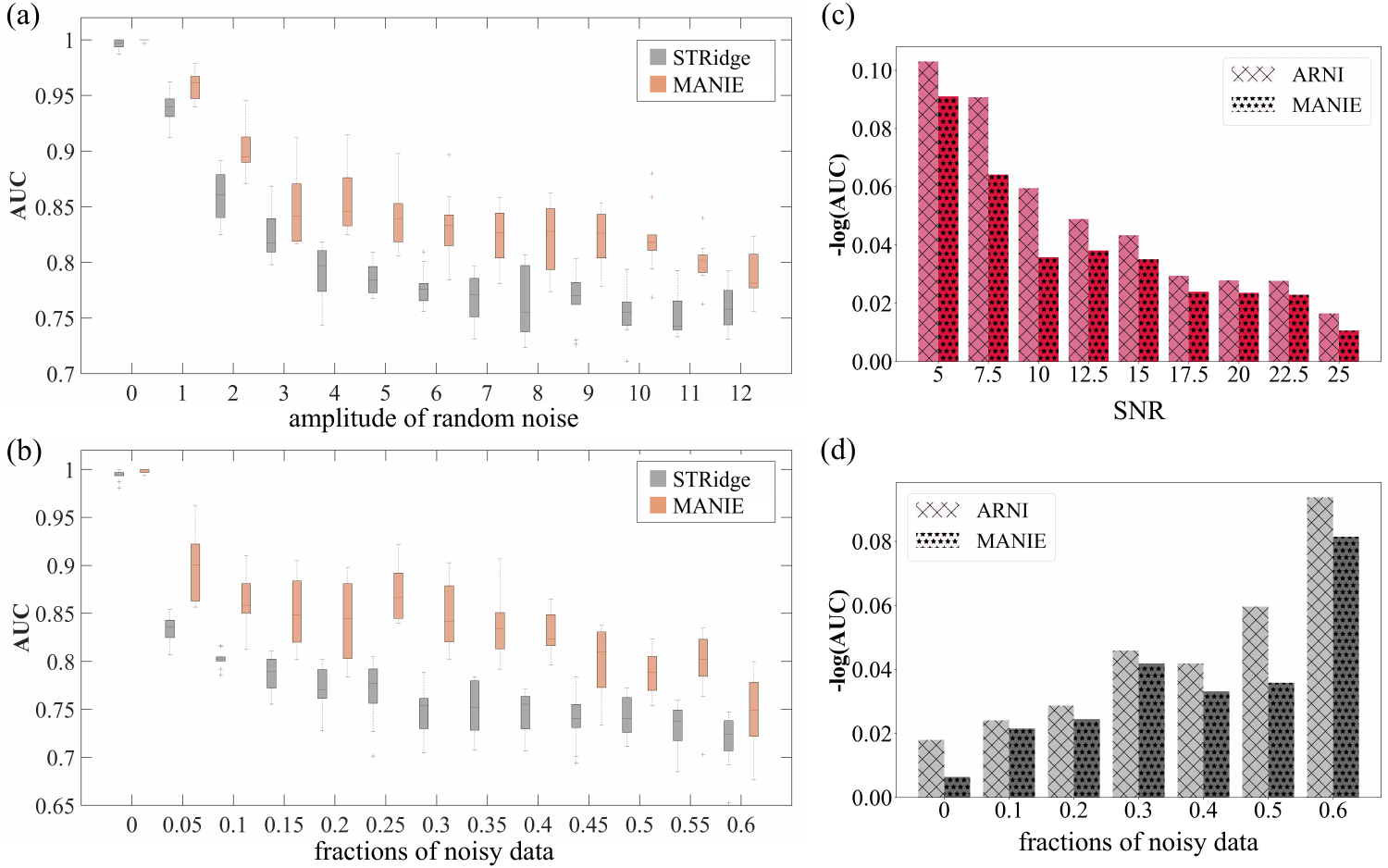}
\caption{demonstrates the influence of noise levels on MANIE in the context of ZK network inference. AUC scores for model-based inference from EG data using MANIE with embedded STRidge, (a) varying by the amplitude of random noise while maintaining a 0.5 fraction of noisy data, and (b) varying by the fraction of noisy data while keeping the amplitude of random noise at 10. AUC scores for model-free network inference, (c) dependent on the signal-to-noise ratio (SNR)/dB and (d) influenced by the fraction of noisy data. The EG process underwent 6 repetitions, each encompassing 10 rounds. Model-free data were aggregated through brief transient time series spanning 5 time steps, originating from 50 distinct initial conditions using a Kuramoto-1 oscillator. 
}
\label{fig3}
\end{figure*}

\textit{Enhancing model-free network inference}. 
Unlike simple model-based network inference methods; we test the efficacy of MANIE in enhancing a more intricate model-free network inference approach, ARNI \cite{16}.
Figure \ref{fig2} shows the substantial performance advancements achieved by MANIE in all scenarios. 
MANIE proves its capacity to augment ARNI's efficacy even under noise-free conditions, although the extent of improvement is relatively lower than that in noisy scenarios. 
Moreover, our observations reveal that reconstructions tend to be superior in scenarios marked by the presence of clean samples.

\begin{figure*}[htbp]
\includegraphics[width=0.85\linewidth]{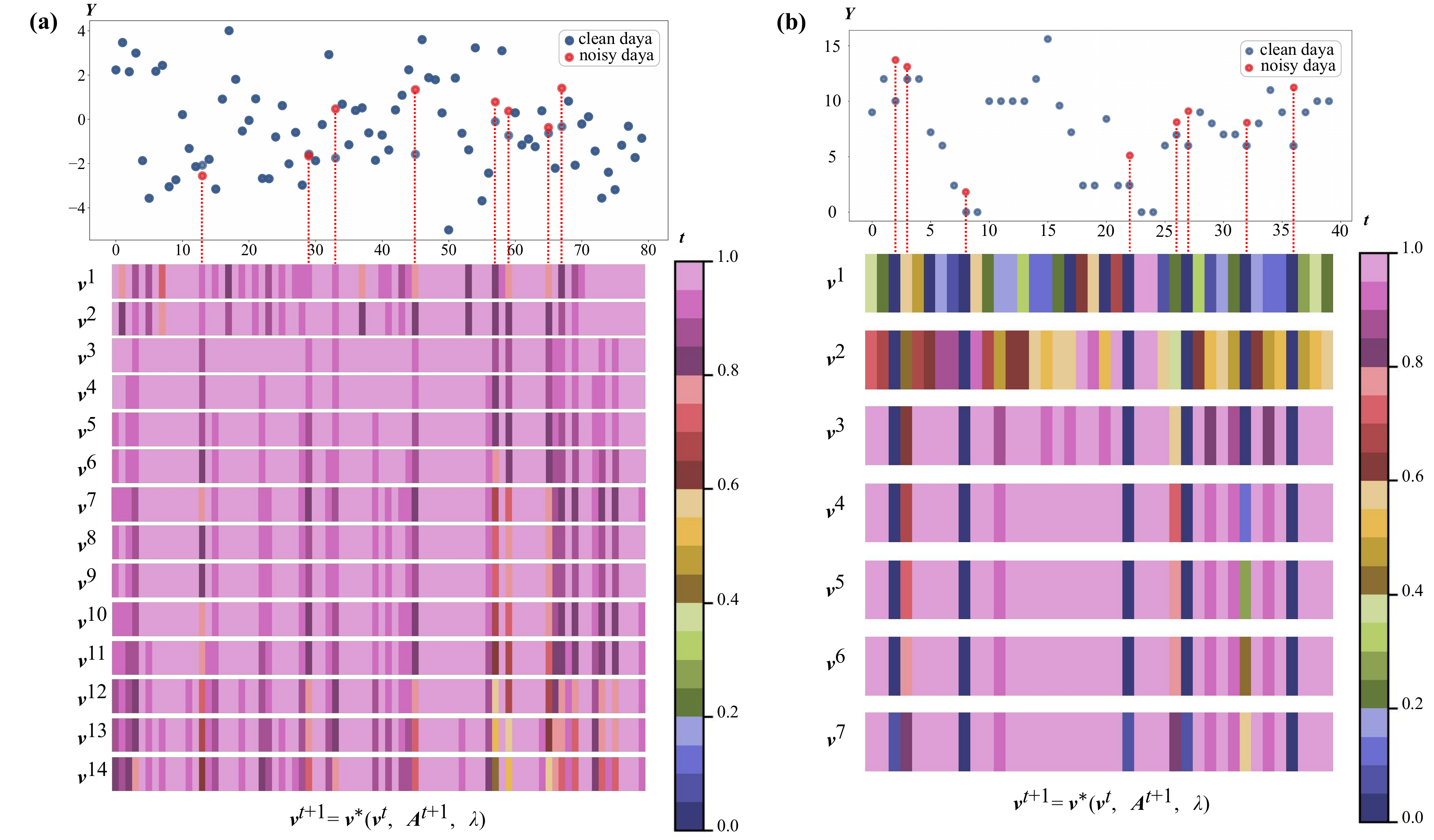}
\caption{illustrates the weight vector $\bm{v}$ optimization. $\bm{Y}$ corresponds to noisy time series. 
Clean data are denoted by blue marks, while red marks indicate data affected by noise. (a) Model-free inference. (b) Model-based inference.}
\label{fig4}
\end{figure*}

Following \cite{19}, we integrate compressed sensing (CS) into MANIE. The experimental outcomes on SIS and CP scenarios are depicted in Figure \ref{fig2}(d)(e). MANIE significantly improves the performance of CS under missing nodes and high noise intensity. Remarkably, it achieves equivalent enhancement even under noise-free conditions. MANIE treats hard-to-fit samples as noisy samples and improves the overall reconstruction performance by reducing the impact of that sample on the loss function. 

\textit{Impact of noise level on MANIE}. 
In Figure \ref{fig3}(a)(c), as noise intensity escalates, the performance of all methods experiences a gradual decline, accentuating the growing effectiveness of MANIE. As illustrated in Figure \ref{fig3}(b)(d), as the proportion of noisy data (noisy data volume/total data volume) increases, the performance of all methods witnesses a diminishing pattern. What remains clear is that, irrespective of fluctuating noise intensities and proportions of noisy data, MANIE consistently demonstrates enhanced noise resistance. Moreover, this valuable attribute persists in both model-free and model-based contexts.

\textit{Visualization of learned $\bm{v}$}. 
As shown in Figure \ref{fig3}, we visualize the learned $\bm{v}$, observing that during the optimization process, the $\bm{v}$ linked to noisy samples tend to decrease. MANIE adeptly alleviates the influence of noisy samples on the embedded network inference methods. Currently, the $\bm{v}$ connected to the majority of clean samples tends to increase, showcasing MANIE's proficiency in safeguarding the contribution of clean samples to the loss function.

MANIE consistently matches or surpasses embedded network inference methods, a conclusion demonstrated in multiple cases. It is notably effective in scenarios featuring clean data samples. Crucially, MANIE exhibits a significant performance improvement even under noise-free and global noise conditions. Our framework's versatility encompasses a wide array of network inference contexts, encompassing model-free and model-based models. It remains straightforward and adaptable, making integration into existing network inference methods seamless. 
However, MANIE's enhancement capacity is confined to certain types of network inference methods. 
A prerequisite for our method's implementation is the availability of sample contributions to the loss function ($\sum_{t=1}^{M}\mathcal{L}\left(\bm{Y}_t,\bm{A}\right)$ or $\sum_{t=1}^{M}\mathcal{L}\left(\bm{Y}_t,\bm{F},\bm{A}\right)$).
In summary, our framework systematically bolsters the noise resistance of model-based and model-free network inference methods, rectifying an overlooked facet. Thanks to its simplicity, adaptability, and efficiency, MANIE has potential for expansive applications.


\nocite{*}


%


\end{document}